\documentstyle[prl,aps,epsfig]{revtex}
\title{Scaling in a continuous time model for biological aging}
\author{R. M. C. de Almeida and G.L. Thomas } 
\address{Instituto de F\'{\i}sica, Universidade Federal do Rio Grande do Sul\\
          Caixa Postal 15051,  91501-970 -- Porto Alegre - RS - Brazil }
\begin{document}
\maketitle
\begin{abstract}
In this paper we consider a generalization to the asexual version of the Penna model for biological 
aging, where we take a continuous time limit. The genotype associated to each individual is an 
interval of real numbers over which Dirac $\delta$--functions are defined, representing genetically 
programmed  diseases to be switched on at defined ages of the individual life.  We discuss two 
different continuous limits for the evolution equation and two different mutation protocols, to be 
implemented during reproduction. Exact stationary solutions are obtained and scaling properties are 
discussed. 
\end{abstract}

\vskip 2cm
\noindent Keywords: Penna model; biological aging \\
\noindent pacs{05.40.+j; 87.10.+e}
\section{Introduction}
A possible explanation for aging in biological populations is based on deleterious mutation 
accumulation 
whose effects are felt at late ages when the intensity of natural selection is lower:  mutations in the 
genome causing diseases genetically programmed to happen at late ages will not prevent reproduction 
and these mutations may spread through the population \cite{char1,rose1}. Population dynamics 
always 
involve non-linear evolution equations and the attractors of these dynamical systems may develop 
correlations in the stationary solutions  that cannot be explained by the analysis of the equations {\it 
per se}. On the other hand, analytical solutions to these equations may be very hard to obtain.

The Penna model for biological aging \cite{penn1} has been proposed as a tool to perform  numerical 
simulations of dynamical systems describing age-structured populations subject to genetic mutations 
and to natural selection,   allowing quantitative analysis of the possible 
attractors. It has been successfully applied to many different aspects of biological aging, considering 
asexual and sexual reproduction,  haploid and diploid populations, different reproduction strategies, 
etc.. For recent reviews see \cite{amer1,moss1}.

The asexual version of the Penna model considers   a bit-string associated to each individual of a given population. 
This bit-string is inherited from the individual's parent and contains information about the genetically programmed 
diseases that the individual will develop during its life. A one in the k$^{th}$ bit of the  string 
implies a disease to be switched on at age k. Each individual dies whenever the $T^{th}$ disease 
happens in its life, that is, the locus of the $T^{th}$ bit  set to one  gives information about the inherited  
maximum age this individual may reach. To avoid an exponential growth of the population, a 
Verhulst factor  is considered through a survival probability given by $( 1 - N(t)/N_{max})$, where 
$N(t)$ is the total population at time $t$ and $N_{max}$ is the carrying capacity of the environment. 
At each time step the living individuals grow older one time unit - usually called year - and the 
corresponding bit in its bit-string is read, testing for the genetically programmed death. After a 
juvenile period, ending at age $R$, the individuals start to reproduce. Different assumptions on the 
fecundity after $R$ have been considered to describe different situations ranging from constant 
fecundity up to age $R_f$, with no reproduction after that,  to increasing with age fecundity, 
describing trees or crustaceans  \cite{moss2} that can grow very old. Also, different protocols have 
been proposed to  
produce the offspring bit-strings. Asexual reproduction is simulated by considering   replicas of the 
parent bit-string except for $M$ bits that may suffer deleterious mutations. Sexual reproduction protocols 
mix the information contained  in both parents  bit-strings and then consider mutations. 
There are many different results, for different parameters and protocols, but  it is clear that in general 
the model grasps the central issue  of decreasing  natural selection intensity with age, yielding  good 
results for the age structure of populations. Analytical approaches to these models have been 
considered for the asexual\cite{deal1} and sexual\cite{deal2} versions of the Penna model, confirming 
simulations results. 
 
Many parameters must be considered in the models. Some of them are inescapable ones: there are 
many different systems and species in nature and if one wishes to describe each of them as a  
consequence of  such a general law as natural selection,  the  model  to be proposed should  allow for 
different sets of parameters, as mutation rate, for example. However, some of the parameters are intrinsic to the numerical techniques used, such as the length $B$ of the bit-strings. These 
parameters may be turned inoffensive if scaling could be found, that is,  they could be related to  
natural units of the system. In a recent paper, Malarz \cite{mala1} considered simulations of the 
Penna model with different bit-strings lengths and concluded that a scaling may not be possible.

When scaling is to be considered, it is always interesting to avoid discrete equations, since the size of 
the time steps relatively to some parameters may be relevant to the scaling. Here we consider the 
asexual version of the Penna model, as  proposed in reference \cite{deal1}, and take its continuous 
time limit. It turns out that age must also be  continuous, and bit-strings turn into strings of 
real numbers of semi-infinite length, over which  a sum of Dirac $\delta$--functions is defined. We 
show 
that the lack of scaling is a consequence of the mutation protocol used, when the mutation probability 
is proportional to the inverse of the bit-string length, $B^{-1}$. On the othar hand, when a Poisson distributed mutation 
probability is considered, as suggested by Bernardes and Stauffer \cite{amer3},  we show that the results are independent of the 
string length considered, provided it is  longer than the maximum possible age predicted for the 
population. We do not contradict previous work without scaling since that work \cite{mala1} used  a 
mutation probability that is proportional to the inverse of the bit-string length. Moreover, besides the scaling with the length of the bit-string, we also show that further scaling properties arise when an adequate continuous time limit is taken to the evolution equations, allowing the definition of $R$ as a natural time unit for each population. 

The paper is organized as follows. In the next section we take the continuum limit of  the discrete 
model for asexual populations by considering  two possible continuous time evolutions. In section 3 we analyze  possible mutation operators and in section 4 
we present solutions and discuss scaling properties. Finally in section 5 we discuss the results and conclude that a full scaling may be found only when adequate mutation protocol and time evolution equations are considered. 
 
\section{From discrete to continuous time}

An analytical approach for the asexual, discrete version of the Penna model \cite{deal1} considers 
relative  populations $ x(a,m,t) $, given as 
\begin{equation}
x(a,m,t) = N(a,m,t)/N_{max} \ ,
\end{equation}
where $N_{max}$ is the carrying capacity of the environment  and $N(a,m,t)$ is the number of 
individuals at time $t$, with age $a$ and death age $m$, 
that is, the age at which the $T^{th}$ bit one of its genome will be read.
The time evolution of these populations is described by the following set of discrete time equations:
\begin{eqnarray}
\label{eqdi}
\begin{array}{ll}
x(a+1,m,t+1)=(1 - x(t))\;\; x(a,m,t)  &  \ \ \ \mbox{      for }0 \; \leq \; a \; < \; m  \ , \\
x(a+1,m,t+1)= 0 &  \ \ \ \mbox{     for } a \; \geq \; m \ . 
\end{array}
\end{eqnarray}
Here $x(t)$ is the total population at time $t$, that is, it is the sum over ages $a$ and death ages $m$ 
of $x(a,m,t)$. The equation for $a=0$ states how offspring is produced. The related equation is
\begin{equation}
x(0,m,t+1) = (1 -x(t)) \; \sum_{a=R} \ \sum_{m'=a+1} F(m,m')\;\;  x(a,m',t) \ ,
\end{equation}
where the birth matrix  $F(m,m')$ is the probability that a parent with death age $m'$ 
gives birth to a child with  death age $m$. When only bad mutations are considered, $F(m,m')$ is 
triangular and it is 
possible to obtain the exact stationary solution for the above evolution equations \cite{deal1}, provided the birth 
matrix fulfills  the following  conditions:
\begin{enumerate}
\item $F(m,m')$ is a triangular matrix such that $F(m, m')=0$ for
$m>m'$, that is, parents cannot give birth to offspring with
larger life expectancy (it corresponds to only bad mutation in the
Penna model). This is not biologically unrealistic for well adapted
populations - advantageous mutations are expected to be extremely
rare due to the large times required  for noticeable species
evolution;

\item $F(m, m) \neq 0$, that is, the probability that the parent
gives birth to offspring with its expected life  length is
different from zero. This condition is also expected in
biological populations, and

\item $F(m,m) <  F(m',m')$ if $m>m'$, that is, the probability that a
parent gives birth to offspring  with its expected life length $m$
decreases with $m$. In other words, the larger the parent expected
life length, the larger the probability that a difference in the
genetic charge of the offspring effectively reduces their expected
life length (the better the genetic code, the larger the number of
events that can spoil it). 
\end{enumerate}
In the original Penna model, mutations are implemented in the following way:  for each newly 
produced bit-string, $M$ loci on the bit-string with $B$ bits  are randomly chosen and set equal to 1, 
regardless their previous state. The birth matrix for the asexual Penna model  can then be explicitly 
obtained \cite{deal1} by considering the probability of one mutation happening in a locus with bit 
zero, and  estimating the change in death age  caused by that mutation. The birth matrix thus obtained 
fulfills the above conditions. 

The results depend on the reproduction strategy. For a given initial reproduction age $R$ and same 
mutation rate $M$,  it can be proven that for final reproduction age $R_f \to \infty$, with constant 
fecundity after $R$, there is a maximum possible life span $m_{max}$ for the stationary state. For initial 
populations 
containing $m>m_{max}$, if $R<m_{max}$ the population evolves towards an age structure that 
gives a survival rate fairly constant up to $R$, and  then decaying  smoothly to zero as $a \to m_{max}$.  If, 
however, $R_f<m_{max}$ or  the initial population  has a maximum life span $m^*$ smaller than 
$m_{max}$,  the survival rate for the stationary solution decreases smoothly up to  $R_f$ or $m^*$, 
where it abruptly decays to 
zero. In the special cases where $R_f=R$ or $R>m_{max}$, reproduction happens  for only one year 
and  the whole population die after it: this is the case of semelparous species that present catastrophic 
senescence \cite{amer1,moss1}.

Here we present a continuous version of this model, and discuss different ways of implementing 
mutations. We start by considering time as a continuous variable, hence age should also be 
continuous. As death can happen at any moment in  time, the genome must be described by a real 
variable interval. To each individual we associate  a genome function $f(\ell)$ of a real 
variable $0\leq \ell < L$, where the string length  $L$ may possibly go to infinity.  This function is a sum of Dirac 
$\delta$--functions, that is
\begin{equation}
f(\ell ) = \sum_{i=1} \delta (\ell  - \ell_i) \ ,
\end{equation}
where the locations of the $\delta$--functions, $\ell_i$, represent the ages at which genetical diseases 
are switched  on.
We  define the health status of an individual as
\begin{equation}
H(a) = \int _0^a f(\ell) d\ell 
\end{equation}
such that  the  death age of the individual is reached when $H(a)=T$, that is, when the $T^{th}$ 
$\delta$-function -- or the $T^{th}$  genetically programmed disease -- happens in its life.
Now  $x(a,m,t) \; da \; dm $ is  the relative number of individuals at time $t$ with age between $a$  and 
$a+da$ and death age between $m$ and $m+dm$.  The total population $x(t)$ is calculated from 
$x(a,m,t)$ as 
\begin{equation}
x(t)=\int_0^{m^*} dm \int_0^m da \;\; x(a,m,t) \ ,
\end{equation}
where $m^*$ is the maximum death age in the population ($m^*\leq m_{max}$). There are two  continuous  time evolution 
equations for  $x(a,m,t)$  that are compatible with the discrete limit Eqs. (\ref{eqdi}), as we will 
discuss now.

\subsection{The Volterra description}

This approach is based on the product integration calculus proposed by Volterra in 1887 
\cite{volt1,volt2}. The first of 
Eqs.(\ref{eqdi}) can be written in the continuous version as
\begin{equation}
\left(\frac{x(a+da,m,t+dt)}{x(a,m,t)}\right) ^{1/(\kappa dt)} = ( 1- x(t)) \ ,
\end{equation}
where the left hand side has  been taken to the power $1/(\kappa dt)$ and $\kappa$ is a constant with  
dimensions of time$^{-1}$ such that $\kappa \Delta t =1$ in the discrete limit: $\kappa^{-1}$ is 
a convenient time unit. The above equation can be rewritten as
\begin{equation}
\label{dlog}
\frac{d}{dt} \ln{x(a,m,t)} = \kappa \ln{(1-x(t))} \ ,
\end{equation}
 where the total derivative can be displayed as 
\begin{equation}
\label{dtotal}
\frac{d}{dt} \ln{x(a,m,t)} = \frac{\partial  \ln{x(a,m,t)} }{\partial a} \frac{da}{dt} +   \frac{\partial  \ln{x(a,m,t)} }{\partial m} \frac{dm}{dt} +    \frac{\partial  \ln{x(a,m,t)} }{\partial t}\ .
\end{equation}
Clearly $da/dt=1$  and $dm/dt=0$. 
The stationary solution ($\frac{\partial  \ln{x(a,m,t)} }{\partial t}=0$)  for Eq.(\ref{dlog}) is then
\begin{equation}
\label{solm}
x(a,m)=x(0,m)(1-x)^{\kappa a } \ ,
\end{equation}
where we have dropped the time index for the stationary solutions. Observe that this is  in the same 
form as the solution for the discrete version of the model \cite{deal1}.

\subsection{The orthodox description}

A more orthodox way of writing the continuous limit for Eqs. (\ref{eqdi}) is to rewrite them as 
follows
\begin{equation}
\frac{ x(a+da,m,t+dt) -x(a,m,t)}{\kappa dt } =  -x(t)\;\; x(a,m,t) \ ,
\end{equation}
where $\kappa$ has the same role as in the Volterra description: a convenient unit such that for the 
discrete limit time interval $\kappa \Delta t =1$.  The differential equation is
\begin{equation}
\frac{d}{dt} x(a,m,t) = - \kappa \;\; x(t) \;\;x(a,m,t) \ ,
\end{equation}
with the total derivative analogously given as in eq. (\ref{dtotal}). The  stationary solution is 
\begin{equation}
\label{sols}
x(a,m,t) = x(0,m) \exp{(- \kappa x a)} \ .
\end{equation}

The difference between stationary solutions Eqs.(\ref{solm}) and (\ref{sols}) is not important when $x 
\sim  0$, since in this case  $(1-x)^{\kappa a} \sim  \exp{(- \kappa x a)}$. On the other hand, observe that the above equations allow $x>1$.  Both solutions depend on  
the newborn populations $x(0,m)$ that in turn depend on the birth matrix $F(m,m')$, that is
\begin{equation}
\label{bir1}
x(0,m) = b \int_R^{m^*} dm' \int_R^{m'} F(m,m')\;\; x(a,m')\;\; da \ ,
\end{equation}
where $b$ is the number of offspring produced per fertile individual per time unit.

 Before proceeding with the solution let us obtain $F(m,m')$.

\section{The birth matrix and  mutation operators}

Here we shall consider two ways of implementing mutations, an analogue to the discrete Penna model 
and  a Poisson distributed probability of happening mutations.

\subsection{ Penna model mutations}
Let us assume that the genome length is $L$. One parameter of the model is the number $M$ of 
loci randomly chosen to assign a new $\delta$--function on the genome. Diversely from the discrete 
model, 
the probability of this happening  on a locus already with a $\delta$--function is vanishingly small. 
Hence, 
the only possibility of the offspring death age being  equal to that of the parent occurs  when no new 
mutation happens before  $ m'$.   The mutation operator $A (m,m')$ for exactly one mutation is given 
as 
\begin{equation}
\label{mu}
A(m,m') =  \frac{L-m}{L}\; \delta(m-m') +  \frac{T}{L} (\frac{m}{m'})^{T-1} \;\;\theta(m'-m) \ ,
\end{equation}
where  $\theta(m'-m)$ is the step function (equals to 1 if $m'>m$ and 0 otherwise) that guarantees 
that 
the child death age is always less or equal to that of the parent. 
The birth matrix for $M$ mutation is
\begin{equation}
\label{fmu}
F(m,m')= (A (m,m'))^{M} \ .
\end{equation}
>From Eqs.(\ref{mu}) and (\ref{fmu}) it is clear  the origin of the lack of scaling: as $L$ is varied, 
the 
mutation probability per unit of genome length varies. And as $T$ appears both as a linear coefficient 
and as an exponent, it is at least not trivial to find sets of parameters that would yield equivalent 
solutions.
On the other hand this mutation protocol is too artificial: certainly there are genomes that do not 
suffer mutations at all as there could be some that suffer more than a fixed number of mutations. A more realistic assumption is to consider a fixed (small) mutation probability per unit of 
genome length, taken from a Poisson distribution, as we consider in what follows.

\subsection{Poisson distributed mutations}

Consider the probability of $n$ mutations happening in a given length $\ell$ of a genome as  given by a 
Poisson distribution
\begin{equation}
P(n,\ell) = \frac{(p\ell )^n \exp{(-p\ell )}}{n!} \ ,
\end{equation}
where $p$ is the probability of occurring one mutation (adding a new $\delta$--function) per unit of 
genome length. To build up the birth matrix $F(m,m')$, there are three possibilities to consider:
{\it i)} if $m>m'$, then $F(m,m')=0$, since we consider only the possibility of adding new
 $\delta$--functions to the genome (deleterious mutations); {\it ii)} if $m=m'$, meaning  that 
no mutation occurs  before $m'$ and the offspring death age is the same as that of its parent, and finally
{\it iii)} if $m<m'$, when new $\delta-$functions are added to the genome before the occurrence of the 
parent $T^{th}$ disease.   Taking all the possibilities into account, the birth matrix can be written as
\begin{eqnarray}
\label{pois}
F(m,m') &=&\delta (m-m')  \;\;\exp{(-pm')} \nonumber \\
              &+& \theta (m'-m) \; \left[ p\;\; \exp{(-pm)} \sum_{k=0}^{T-1} C_{T-1}^{k}\;\; (\frac{m}{m'})^k \;\;
(\frac{m'-m}{m'})^{T-1-k}\;\; \frac{(pm)^{T-1-k}}{(T-1-k)!} \right. \nonumber \\
              &+&\left. \frac{T-1}{m'} \;\; \exp{(-pm)} \sum_{k=0}^{T-2} C_{T-2}^{k} 
(\frac{m}{m'})^k \;\;
(\frac{ m'-m}{m'})^{T-2-k} \;\;\frac{(pm)^{T-1-k}}{(T-1-k)!} \right]  \ ,
\end{eqnarray}
where $\theta(m'-m)$ guarantees that $F(m,m')=0$ if $m>m'$ and $ C^k_T \equiv T ! / ( k! \  (T-k)! ) $. Observe that mutational meltdown may be prevented since there is a  non-zero probability that some offspring are bred without additional harmful mutations ($m=m'$) \cite{amer3}. It can be 
shown that the  above birth matrix obeys the normalization condition
\begin{equation}
\int_0^{m'} dm \;\; F(m,m')=1 \ ,
\end{equation}
that is the sum over all possibilities for the offspring genome. Comparing Eqs. (\ref{mu}) and 
(\ref{pois}), we observe that the mutation controlling parameters in the different mutation protocols are respectively $L$ and $p$, that is, 
using Poisson protocol the mutation probability decouples from the genome length $L$. Hence, since $L$ no longer appears in the mutation operator, the results do not depend on $L$, and scaling should be expected. When only bad mutations are considered, there may exist a maximum life span $m_{max}$ to the steady state solutions that is independent of $L$. In this case, the stationary solutions are not affected by bits located after $m_{max}$ and consequently are independent of $L$, provided the mutation rate per genome length does not depend on $L$. In other words, the dependence on $L$  
seems to be responsible for the lack of scaling in Eq.(\ref{mu}) when mutations are implemented as in the discrete version of the Penna model. 

On the other hand, further  scaling properties can be 
found in the solutions of the evolution equations, provided  the Poisson mutation protocol is used, as we show in the next section. 

\section{Solutions and scaling}
To have further insight of what happens we shall consider explicit solutions to the evolution 
equations. From now on we shall consider only Poisson mutation protocol. We first rewrite Eq.(\ref{bir1}) for the newborn offspring using Eq.(\ref{sols}) for the orthodox evolution:
\begin{equation}
\label{bir2}
x(0,m) = \frac{b}{x \kappa} \int_m^{\infty} dm' \;\;[ \exp{(-x \kappa R)} - \exp{(-x \kappa m')}] \;\; 
F(m,m') \;\;
x(0,m')  \ .
\end{equation}
We observe next that  $x(a,m)$ has  dimensions of time$^{-2}$, $b$, $p$, and $F(m,m')$  have 
dimensions of time$^{-1}$,  while $m$, $R$, and $a$ have dimensions of time$^1$ and  $x$ is 
dimensionless. We can take $\kappa^{-1}$ as the time unit and rewrite all equations in dimensionless 
variables, which is equivalent to rewrite all equations setting $\kappa =1$ and reading all quantities 
as 
given in time unities of $\kappa^{-1}$.  We then have one less parameter, and we remark that reading 
$p$ in  units of $\kappa$ links the mutation rate $p$ to the exponential constant of time evolution.

Even with one less parameter, Eq.(\ref{bir2}) is an integral equation, involving all newborn 
population and the birth matrix.  We can solve this equation because $F(m,m')$ is triangular, but we 
must find out  the maximum possible value $m_{max}$ for the death age that any stationary solution 
may present. We do this in an analogous fashion to that used for the discrete version of the model 
\cite{deal1}. We assume that there is a maximum death age $m^*$ in the population. As $F(m,m')$ is 
triangular - there are only bad mutations -  as time proceeds this maximum value cannot increase, 
although too long living individuals may disappear. Offspring with death age equal to $m^*$ can be 
produced only by $m^*$--parents. Using Eq.(\ref{pois}), the newborn equation for $m=m^*$ can be 
written as 
\begin{equation}
\label{msta}
x(0,m^*) = \frac{b}{x} [ \exp{(-xR)} - \exp{(-xm^*)} ] \;\; \exp{(-pm^*)} \;\;  x(0,m^*) \ .
\end{equation}
The trivial solution, $x(0,m^*)=0$, is always possible. A nontrivial solution, 
$x(0,m^*) \neq 0$, is possible if    two conditions are met. First, 
\begin{equation}
\label{cond1}
\frac{b}{x} \exp{(-pm^*)} [\exp{(-xR)} - \exp{(-xm^*)} ] =1 \ ,
\end{equation}
what guarantees that Eq.(\ref{msta}) is satisfied, and a second condition that guarantees that 
$x(0,m^*-dm)\geq0$, 
\begin{equation}
\label{cond2}
\frac{d}{dm} \exp{(-pm)} [ \exp{(-xR)} - \exp{(-xm)}] |_{m=m^*} \geq 0 \ ,
\end{equation}
resulting in
\begin{equation}
\label{maxo}
m^* \leq  m_{max} = R + \frac{1}{x} \ln{\left(\frac{p+x}{p}\right)} \ .
\end{equation}
All these calculations may be repeated for the Volterra evolution equations. The equivalent equations 
to be solved have the same form as Eqs.(\ref{cond1}) and (\ref{cond2}) with $x$ replaced by $-
\ln{(1-x)}$, and the  corresponding limit for the maximum life span in a given population is
\begin{equation}
\label{maxp}
m^* \leq  m_{max} = R - \frac{1}{\ln(1-x)} \ln{\left(\frac{p-\ln(1-x)}{p}\right)} \ .
\end{equation}
A limit for the life span of a population  means that if, at initial times, the maximum life span  in the 
population is $m^*\geq m_{max}$, in the stationary solution the population goes to a state where 
$m^*=m_{max}$. But if  initially $m^* < m_{max}$, the population remains with that maximum life 
span, since in this model only bad mutations are allowed to happen. For each possible $m^*$ ($ \leq m_{max}$) the total population $x$ may be obtained from Eq.(\ref{cond1}) or from its analogue 
for respectively orthodox or Volterra evolutions. $m_{max}$ is the largest  value of $m^*$ that 
satisfies  both conditions (\ref{cond1}) and (\ref{cond2}) or  their analogues.

These two limits for the maximum life span for different evolution protocols depend differently on the value of the total population.
We can observe some scaling properties regarding the maximum possible life span $m_{max}$ and 
total population $x$. If we have, for every $\alpha > 0$,
\begin{eqnarray}
\label{scale1}
p'&=& \alpha \; p \ , \nonumber \\
R'&=& \frac{R}{\alpha} \ ,  \\
b'&=& \alpha \; b \ . \nonumber
\end{eqnarray}
Then for the orthodox evolution 
\begin{eqnarray}
\label{scale2}
x'&=& \alpha \; x \ , \nonumber \\
m_{max}'&=& \frac{m_{max}}{\alpha} \ , \\
x'( a/\alpha , m/\alpha)&=& \alpha^3 x (a,m) \ . \nonumber
\end{eqnarray}
On the other hand, for the Volterra description the  total population scaling is different:
\begin{eqnarray}
x'&=&  1-\; (1-x)^{\alpha} \ , \nonumber \\
m_{max}'&=& \frac{m_{max}}{\alpha} \ , 
\end{eqnarray}
and the relation between  population densities $x(a,m)$ and $x'(a/\alpha, m/\alpha)$ is far from 
trivial:
\begin{equation}
\left(1-\int_0^{m^*} dm \int_0^m da \;\; x(a,m)\right)^{\alpha} = 1-\int_0^{m^*/\alpha} d(m/\alpha) 
\int_0^{m/\alpha} d(a/\alpha) \;\; x'(a/\alpha, m/\alpha) \ .
\end{equation} 

We note that for both evolutions the value of maximum life span does not depend on $T$. Hence 
varying the number 
of harmful diseases that the individuals may tolerate  will not change the maximum life span, although it can 
change the survival rate for $a < m_{max}$. The reason is that both conditions 
stated in Eqs. (\ref{cond1}) and (\ref{cond2}) are generated by the first term in the expression for the 
birth matrix, Eq.(\ref{pois}), that gives the probability for  a child  with the same death age as 
its parent. This term depends only on the parent death age, and not on the number of diseases that 
there are in the parent genome before $m'$. This is also in agreement with the fact that the 
increasingly better medical care  has  increased human life expectancy but has not equally changed the 
maximum life span \cite{demo,azbe}. In the Penna model the effect of proper medical care is 
equivalent to a shift of the maximum number $T$ of genetically programmed diseases that an 
individual may endure\cite{niew}. In Fig.(\ref{mmaxvsp}) the plots of maximum life span and total 
population as functions of mutation probability $p$ for different values of fecundity $b$. The 
maximum life span decreases with increasing $p$ and decreasing $b$.  

\parbox[t]{8.2cm}
{
\begin{center}
\epsfig{figure=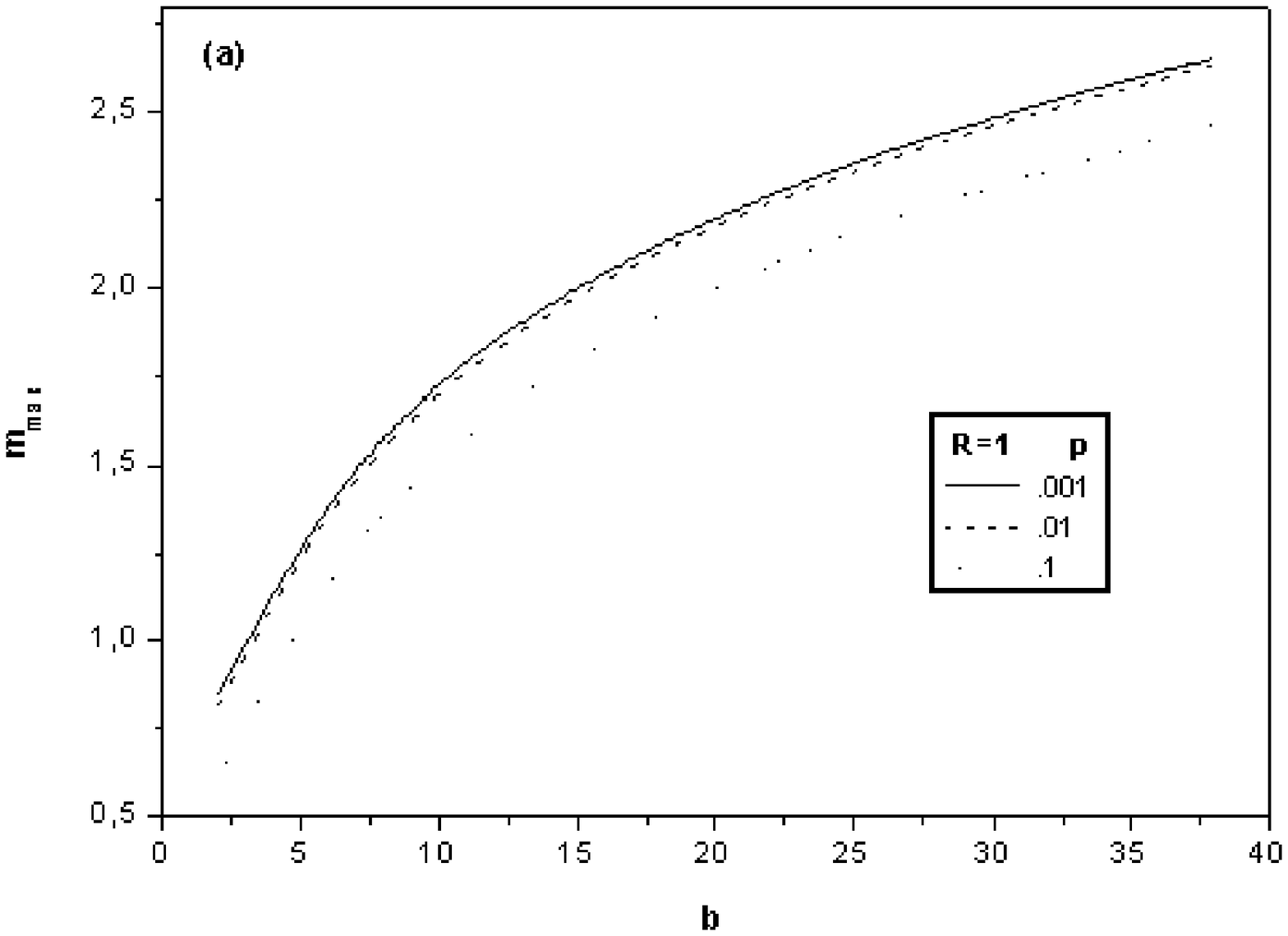,width=9cm,height=5.45cm} 
\end{center}
} \ \hspace*{.15cm}\ 
\parbox[t]{8.2cm}{
\begin{center}
\epsfig{figure=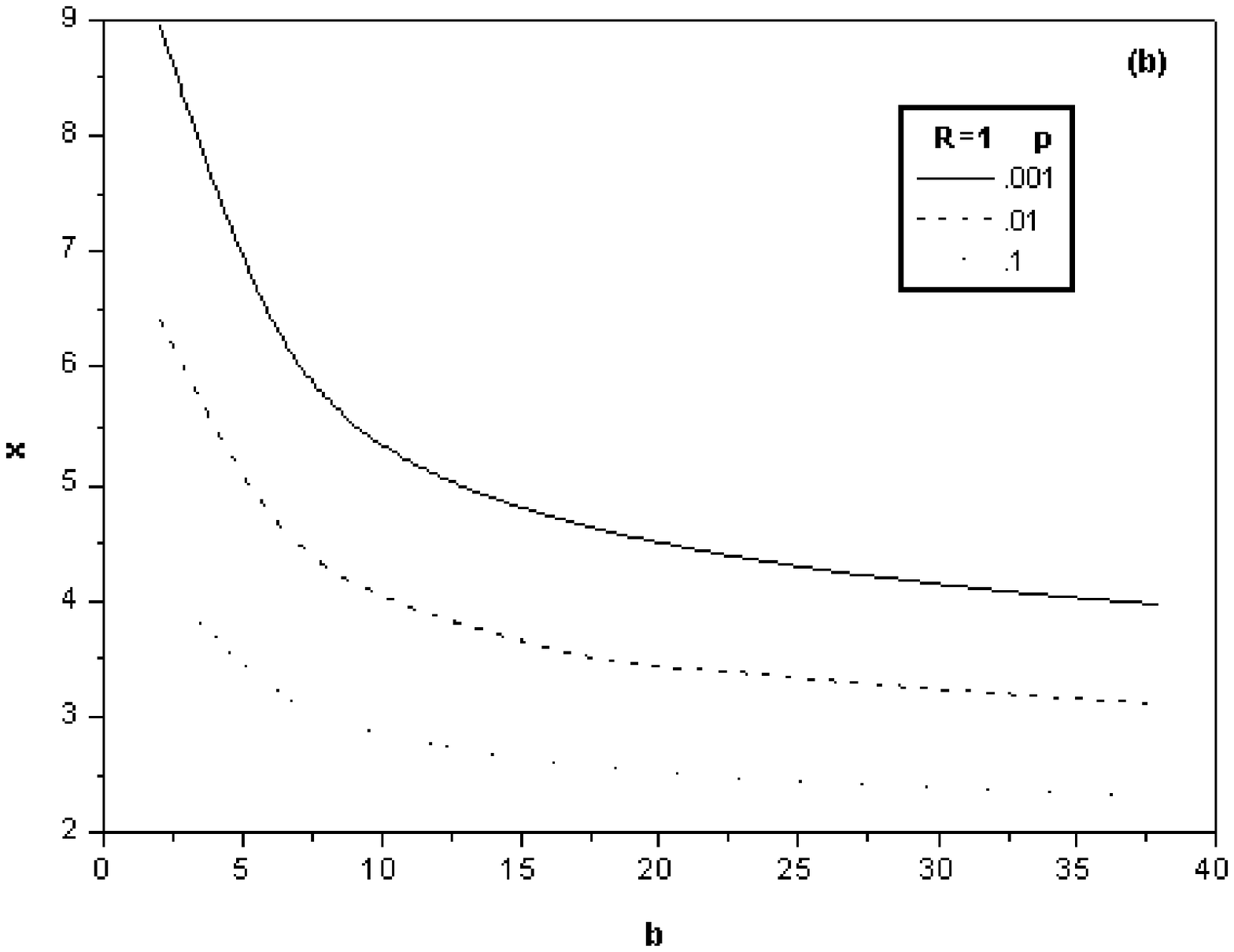,width=8.5cm,height=5.5cm} 
\end{center}
}
\vspace*{-.5cm}
\begin{figure}
\caption{Plots of (a) maximum life span and (b) total population as functions of fecundity $b$ for 
different values of mutation probability $p$. The maximum life span decreases with increasing $p$ and decreasing $b$. Observe that here $x$ can be greater than one, since we are considering the orthodox time evolution.}
\label{mmaxvsp} 
\end{figure}
 
To fully appreciate the scaling present in the solutions we must solve Eq.(\ref{bir2}). For that we 
first transform the integral equation into a differential one. Consider  a stationary population 
with $m^*=m_{max}$, then for $m>R$ the birth equation can be rewritten as 
\begin{eqnarray}
\label{bir3}
\lefteqn{ \left( 1 - b \exp{(-pm)} \; \frac{ \exp{(-xR)}- \exp{(-xm)}}{x} \right) x(0,m) =} \nonumber \\  & & 
b \;p\; \exp{(-pm)} 
(\frac{m}{m'})^{T-1} \int_{m}^{m_{max}}
\; \frac{ \exp{(-xR)} - \exp{(-xm')}}{x}\;\; f(m'-m) \;\; x(0,m') \;\; dm' \ ,
\end{eqnarray}
where $f(m'-m)$ is defined as 
\begin{equation}
f(m'-m)= \sum_{k=0}^{T-1} C_{T-1}^{k} p^k \frac{(m'-m)^k}{k!} 
+ \sum_{k=0}^{T-2} C_{T-1}^{k+1} p^k \frac{(m'-m)^k}{k!} 
\end{equation}
such that, after some calculations we can write
\begin{equation}
F(m,m') = \exp{(-pm)} \;\; \delta(m'-m) + p\;\exp{(-pm)}\;(\frac{m}{m'})^{T-1} \;f(m'-m) \; 
\theta(m'-m) \ .
\end{equation}
We now rewrite Eq.(\ref{bir3})  as
\begin{eqnarray}
\lefteqn{ \frac{\exp{(-pm)}}{b \; m^{T-1}}\left( 1 - b \exp{(-pm)} \;\; \frac{ \exp{(-xR)}- \exp{(-
xm)}}{x} \right) 
x(0,m) = } \nonumber \\ & & \ \ \ \ \ \ \ \ \ \ \ \ \ \ \ \ \ \ \ \ \ \ \int_{m}^{m_{max}}
\frac{ \exp{(-xR)} - \exp{(-xm')}}{xm'^{\;T-1}}\;\; f(m'-m) \;\; x(0,m') \;\; dm' \ .
\end{eqnarray}
We then differentiate $T$ times both sides, and use the fact that we can calculate the value of $f(m'-
m)$ and all its derivatives at $m=m'$. We then arrive to the following differential equation:
\begin{equation}
\label{eqpar}
\frac{\partial ^T}{\partial m^T} \left( \frac{\exp{(pm)}}{bm^{T-1}}x(0,m) \right) = \exp{(pm)}
\frac{\partial ^T}{\partial m^T}  \left(\frac{\exp{(-xR)}- \exp{(-xm)}}{xm^{T-1}}\;\;\exp{(-pm)} \;\; 
x(0,m) \; .
\right)
\end{equation}
This equation may be numerically solved.  To obtain  the analogue of the above equation for the 
Volterra evolution it suffices to replace $x$ by $-\ln{(1-x)}$. Nevertheless, in what follows we will 
consider only the orthodox evolution.

\noindent\parbox[t]{8.3cm}
{ 
We have numerically solved Eqs.(\ref{eqpar}) for $T=1$ and $T=2$. Due to the scaling properties we 
can use the initial reproduction age $R$ as our `natural' unit. In Fig.(2) where 
we plot $x(0,m)/x(0,R)$ for  sets of $R$, $p$ and $b$ respecting the scaling conditions given 
in Eqs.(\ref{scale1}), for both $T=1$ and $2$. The scaling is clearly shown. In what follows we shall use $R=1$ and mutation and birth rates $p$ and $b$ are given in units of $R^{-1}$.

\hspace*{1em} The effect of  different mutation and birth rates on the population can be monitored through the   
survival rate $S(a)$  that is  defined as 
\begin{eqnarray}
S(a) &= & \left( \frac{x(a+da)}{x(a)}\right)^{1/da} \nonumber \\
&=&  \exp{(\frac{d \ln(x(a))}{da})} \ .
\end{eqnarray}
} \ \hspace*{.15cm}  \ 
\parbox[t]{8.5cm}{
\vspace*{-.5cm}
\begin{center}
\psfig{figure=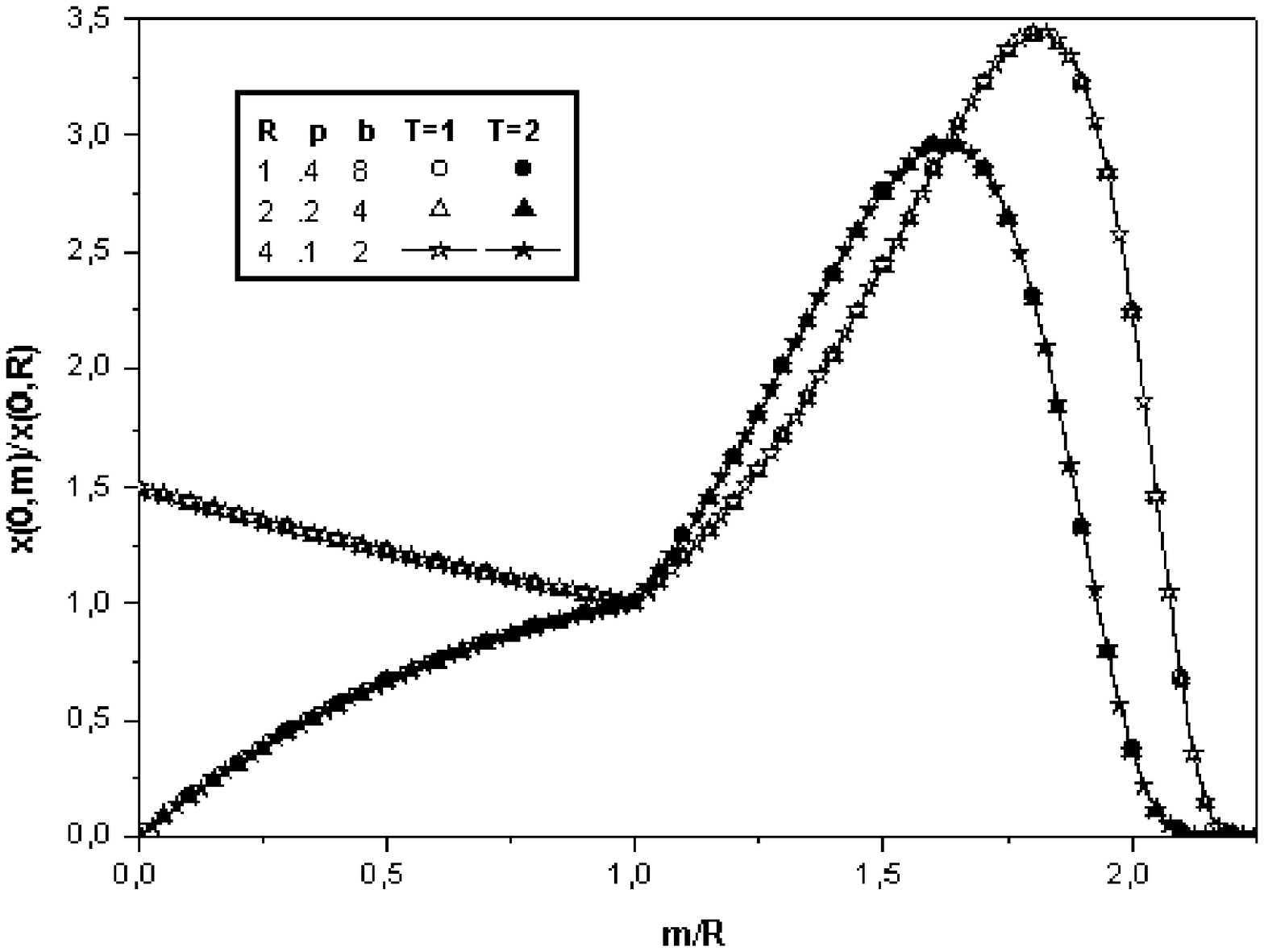,width=8.5cm,height=5.5cm} 
\end{center}
\vspace*{-.6cm}
{\small FIG. 2. Plots of  $x(0,m)/x(0,R)$ for  sets of $R$, $p$ and $b$ respecting the scaling conditions, as given in the inset, for both $T=1$ and $2$. This behaviour refers to the orthodox evolution with Poisson mutation protocol. }
}
\addtocounter{figure}{1}

\noindent\parbox[t]{8.3cm}
{ 
\hspace*{1em} Figure (3) shows how the survival rate changes when the fecundity $b$ varies for the same 
value of mutation rate $p=0.001$. As $b$ increases, the total population $x$ also increases and the 
survival probability is uniformly  reduced at all ages. Also, the maximum life span and the onset 
of senescence, marked by the  strong reduction in survival rates, are shifted to earlier ages.  The figure shows data for both $T=1$ and $2$. The maximum life span $m_{max}$ remains the same for different $T$, but one can see that  at the same age, higher values of $T$ imply lower survival 
rates, that is, better medical care  decreases the strength of natural selection such that more mutations accumulate at later ages.

\hspace*{1em} Figure (4) shows how the survival rate depends on the mutation rate $p$, for the same 
value of $b=4$.  As expected,  a smaller $p$  enhances the maximum life span. However, young individuals present lower  survival rates due to larger total population $x$, that reduces mortality through the exponential pre-factor $\exp(-xa)$.
\addtocounter{figure}{1}

\hspace*{1em} Finally,  Fig.(5) presents the normalized mortality function  $q(a)$, defined as 
\begin{equation}
q(a)= - \ln{\left[\frac{S(a)}{S(0)}\right] } \ ,
\end{equation}
for  different values of $b$, $p$ and $T$. A Gompertz region is clearly seen, and also a deviation 
from the exponential behaviour for young ages.  
} \ \hspace*{.15cm} \ 
\parbox[t]{8.5cm}{
\vspace*{-.5cm}
\begin{center}
\psfig{figure=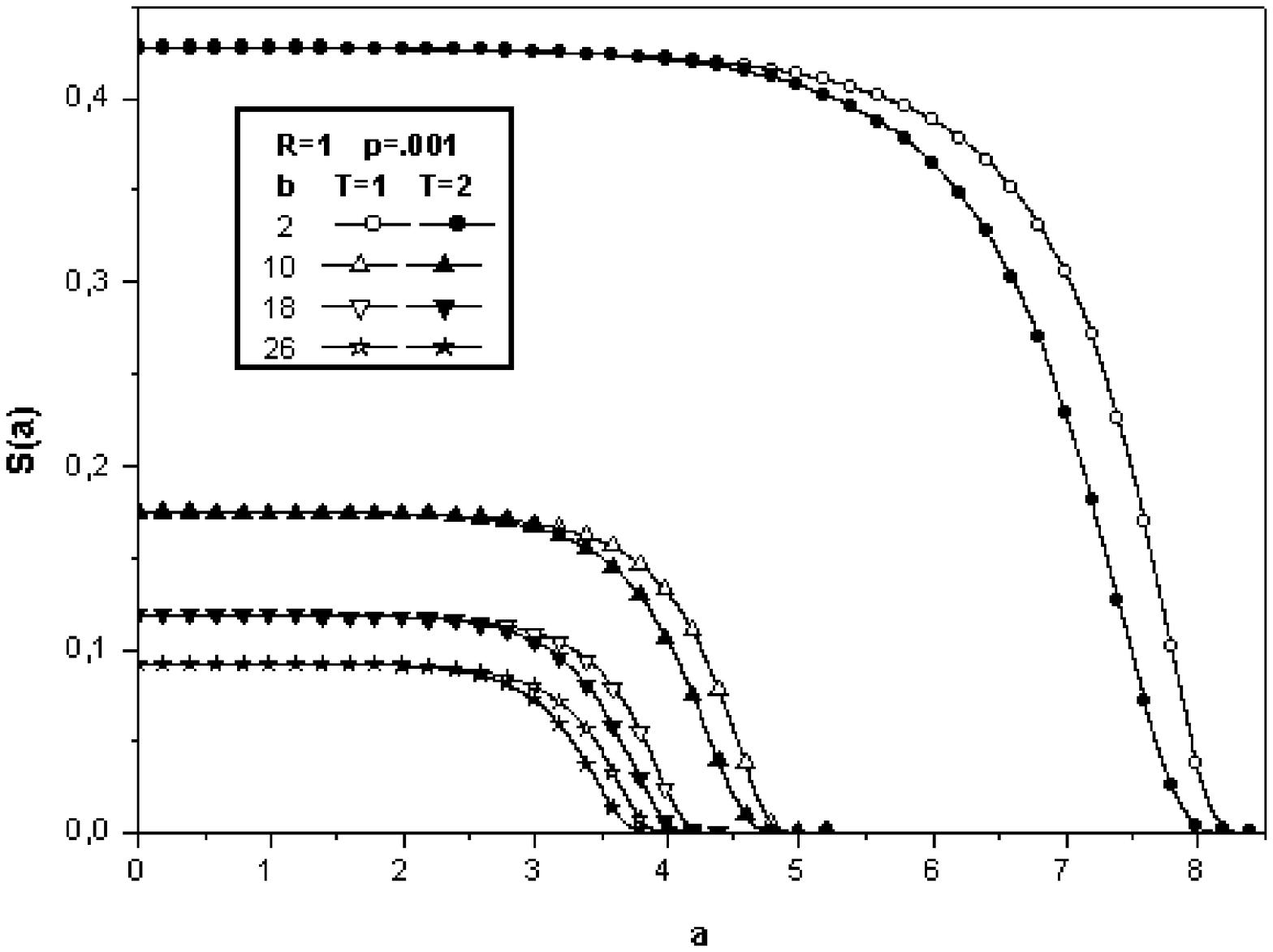,width=8.5cm,height=5.5cm} 
\end{center}
\vspace*{-.6cm}
{\small FIG. 3. Survival rate $S(a)$ versus $b$ for mutation rate $p=0.001$ for the orthodox evolution with Poisson mutation protocol. Observe that smaller $T$ imply higher $S(a)$ for the same $a$. }
\vspace*{.8cm}
\begin{center}
\vspace*{-.7cm}
\epsfig{figure=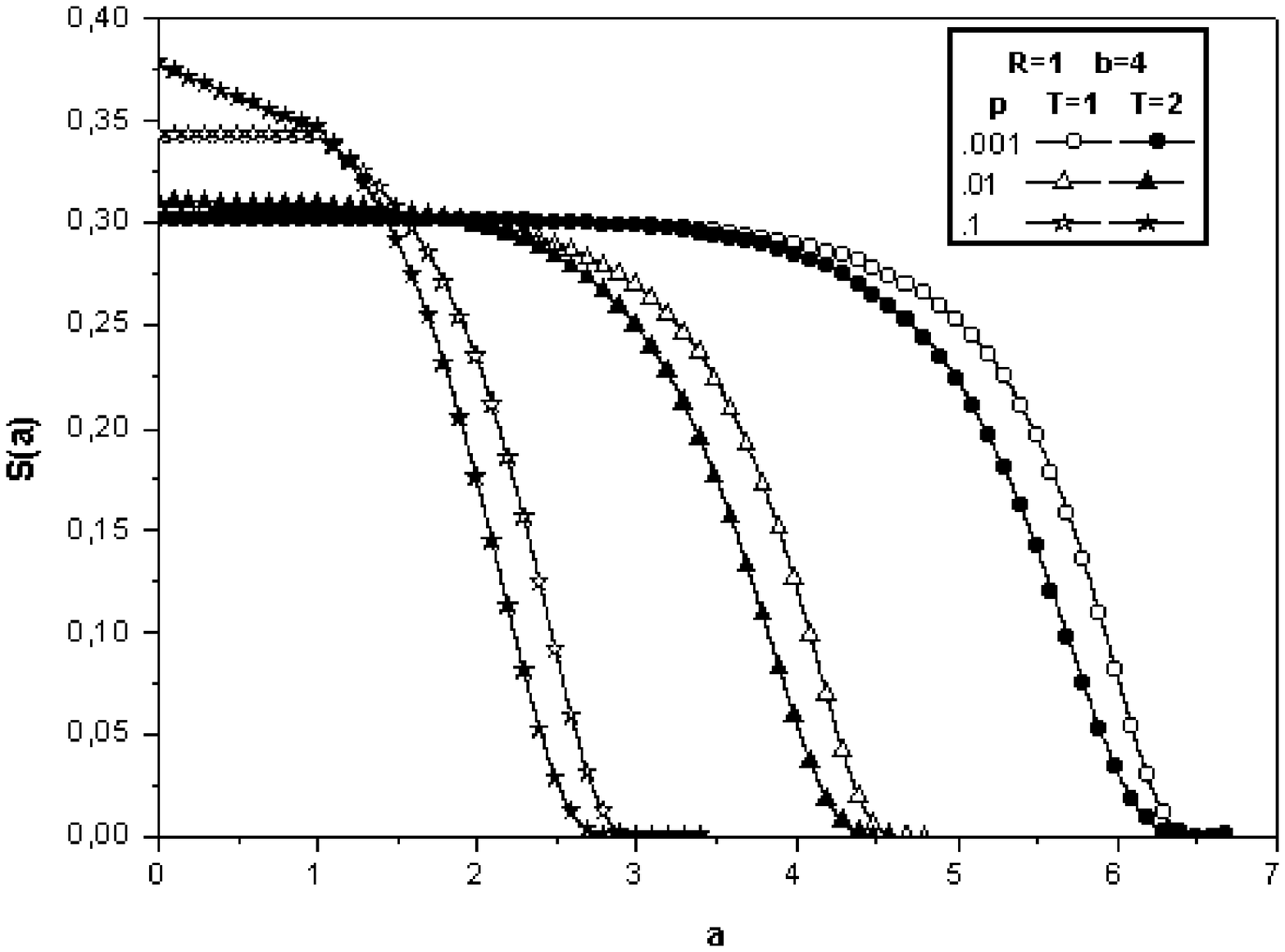,width=8.5cm,height=5.5cm} 
\end{center}
\vspace*{-.8cm}
{\small FIG. 4. Survival rate $S(a)$ versus $p$ for birth rate $b=4$ for the orthodox evolution with Poisson mutation protocol.}
}

\noindent\parbox[t]{8.2cm}
{
\section{Conclusion and discussion}

\hspace*{1em} In this paper we have generalized to continuous time the discrete asexual Penna model for aging, such 
that the genome of the individuals is now represented by a sum of $\delta$-functions on  continuous 
interval, $\delta(a-a_0)$, each one representing a disease to be switched  on at the age $a_0$.

\hspace*{1em} Two different time evolution protocols  could be used: Volterra and orthodox time evolution equations 
are compatible with the discrete formulation of the Penna model. However, we showed that a complete 
scaling is only found with the orthodox evolution equations (eqs. (\ref{scale1})-(\ref{scale2}) ). 

\hspace*{1em} Also, two different protocols may be considered  to describe mutations happening in offspring genomes. One  considers a mutation rate that is proportional to the length of the genome. This is equivalent to the mutation protocol  commonly used with the discrete
} \ \hspace*{.15cm} \ 
\parbox[t]{8.2cm}{
\begin{center}
\vspace*{.5cm}
\epsfig{figure=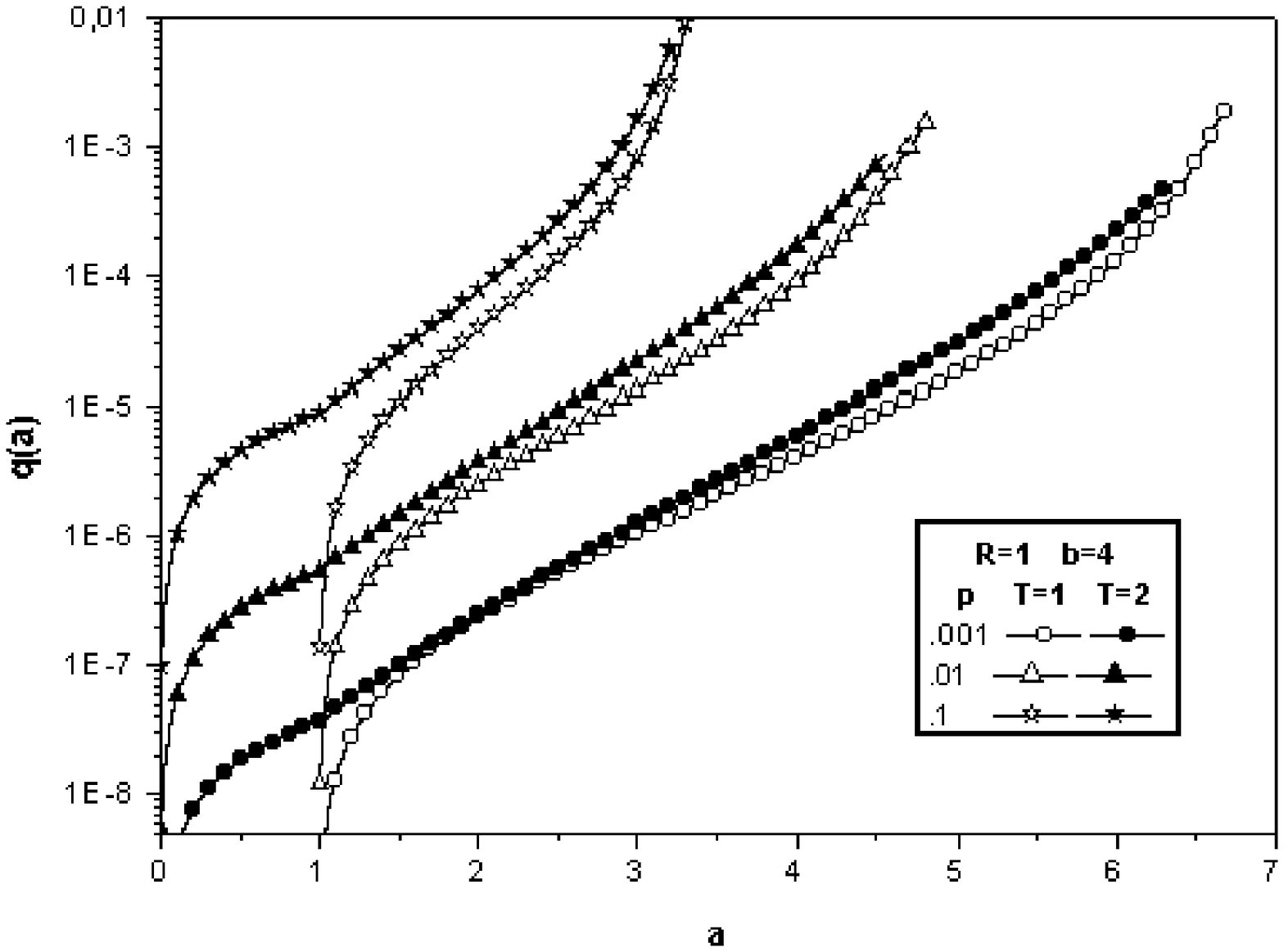,width=8.5cm,height=5.5cm} 
\end{center}
\vspace*{-.8cm}
{\small FIG. 5. Plots of (a) maximum life span and (b) total population as functions of fecundity $b$ for 
different values of mutation probability $p$. The maximum life span decreases with increasing $p$ and decreasing $b$. Observe that here $x$ can be greater than one, since we are considering the orthodox time evolution.}
}

\vspace{.1cm}\noindent version of the Penna model, where scaling may not be found due to the somewhat artificial  dependence of the mutation rate on the length of the genome.  A second mutation protocol considers a Poisson probability distribution, with mutations occurring with a fixed probability $p$ per  unit of genome length. As this last mutation protocol   does not depend on the 
genome length, the whole set of stationary solutions does not depend on $L$.   

Morever, in the special case of the orthodox time evolution and Poisson mutation protocol, we  could find further  scaling relations such that   the initial reproduction age 
$R$ may be  taken as the natural time unit of the system. In this sense, all rates, as mutation and birth rates, are 
renormalized in units of $R^{-1}$.
One direct consequence of this scaling is the correlation between earlier initial reproduction age and 
earlier senescence, an effect that is due to the scaling inherent to the model that shows up in the stationary solutions 
of the  population dynamics equations. That is, here demographic effects only are responsible for this correlation and  antagonistic pleiotropy effects are not needed.

The generalization of sexual version of the Penna model is now under investigation and shall appear 
elsewhere.

\section*{Acknowledgement}
The authors thank D. Stauffer for suggesting this work and fruitful discussions with S. Moss de Oliveira, T.J.P. Penna, and A.T. Bernardes.  This work has been partially finaced by Brazilian agencies CNPq, CAPES and FAPERGS.

\end{document}